\RequirePackage{fix-cm}
\documentclass[smallcondensed]{svjour3}     
\smartqed  
\usepackage{graphicx}
\usepackage[intlimits]{amsmath}   
\usepackage{amssymb}   
\usepackage{natbib}    
\usepackage{hyperref}   
\usepackage[english]{babel}
\usepackage{mathtools}
\newcommand{\bfl}{\mathbf}
\newcommand{\bfg}{\boldsymbol}
\begin{document}

\title{Effects on satellite orbits in the gravitational field of an axisymmetric central body with a mass monopole and arbitrary spin multipole moments}

\titlerunning{Effects on satellite orbits in the gravitational field of an axisymmetric central body}      

\author{Jan Meichsner         \and
        Michael H. Soffel 
}

\authorrunning{J. Meichsner, M.H. Soffel} 

\institute{ J. Meichsner \at
			  Institute for Theoretical Physics, Dresden University of Technology, BZW, Zellescher Weg 17,  01069 Dresden, Germany\\
			  Tel.: +49 (0) 351 463-32379\\
			  Fax: +49 (0) 351 463-37019\\
			  \email{Jan.Meichsner@tu-dresden.de} 
			\and
			M.H. Soffel \at
              Department of Astronomy, Dresden University of Technology, Beyer-Bau, George-B\"ahr-Stra\ss{}e 1,  01069 Dresden, Germany\\
              Tel.: +49 (0) 351 463-34200\\
              Fax: +49 (0) 351 463-37019\\
              \email{Michael.Soffel@tu-dresden.de}  
}

\date{This is an online version of this article published in \\ 
\textit{Celestial Mechanics and Dynamical Astronomy}. \\ 
The final publication is available at \\ 
\url{http://link.springer.com/article/10.1007/s10569-015-9626-3}}

\maketitle

\begin{abstract}
Perturbations of satellite orbits in the gravitational field of a body with a mass monopole and arbitrary spin multipole moments are considered for an axisymmetric and stationary situation. Periodic and secular effects caused by the central gravitomagnetic field are derived by a first order perturbation theory. For a central spin-dipole field these results reduce to the well known Lense-Thirring effects.

\end{abstract}

\section{Introduction}
\label{intro}
As is well known Einstein's theory of gravity leads to so-called
gravitomagnetic effects, i.e., gravitational effects caused by moving or rotating masses. Such effects, e.g., related with a rotating Earth, have been studied intensively in the past. To be mentioned here is the Lense-Thirring effect that describes perturbations of a satellite orbit in the gravitomagnetic dipole field of a rotating body. It was first described by Lense and Thirring (1918, see \citet{lense:1984} for a translation) after extensive groundwork by Einstein. \citet{ciufolini:2004} found the effect to be confirmed within a 10 \% accuracy by a detailed analysis of the orbits of the two LAGEOS satellites. The results, however, are still subject to an ongoing debate (see \citet{iorio2011}). 

Closely related to the Lense-Thirring effect is the precession a torque-free gyroscope experiences due to gravitomagnetism. This Pugh-Schiff effect was proposed to be an alternative test for frame-dragging (see \citet{pugh1959} and \citet{schiff1960}). Although unexpected problems arose \citet{everitt:2011} measured the effect within an accuracy of 13 \%.  

The above mentioned effects solely take the spin dipole into account. Naturally one is also interested in the influence of higher spin multipole moments. Such higher multipole moments come into play via multipole expansions of the so-called gravitoelectric potential $\phi$ (a generalisation of the Newtonian potential $U$ which is often also denoted by $w$) and the gravitomagnetic vector potential $\mathbf{w}$. Both are used in order to parametrise the metric tensor in the first post-Newtonian approximation. For more details on the origin and form of the metric tensor in the first post-Newtonian approximation please see \citet{soffel2003} and \citet{soffeldamourxu:1991}. 

In this paper $\mathbf{w}$ will be our primary concern. As for the gravitoelectric potential $\phi$ we will just use a post-Newtonian mass monopole, i.e., $\phi=GM/r$, where $G$ is the Gravitational constant, $M$ the Blanchet-Damour mass of the central body and $r = \| \mathbf{x} \|$ the distance to its centre of mass. The gravitomagnetic potential $\mathbf{w}$ is induced by a matter current density $\bfg{\sigma}$ ($\sigma^k := T^{0k}/c$, where $T^{\mu\nu}$ is the body's energy-momentum tensor and $c$ the speed of light). In the stationary case and outside a coordinate sphere ${\cal S}$ that fully covers the energy-momentum tensor of the central body $\mathbf{w}$ admits a multipole expansion of the form (e.g. \citet{blanchet:1989:377}) 
\begin{equation} 
\bfl{w} = w_k \bfl{e}_k = - G \sum_{l=1}^{\infty} \frac{(-1)^l l}{l!(l+1)} \epsilon_{kab} J_{bL-1} \partial_{aL-1} \left(\frac{1}{r} \right) \bfl{e}_k \label{vectorpotential}
\end{equation}

with the spin multipole moments
\begin{equation}
	J_L \coloneqq  \int_{\mathbb{R}^3} \epsilon_{ab < k_l} \hat{x}_{L-1 > a} \sigma_{b} d^3x. 
\end{equation}
In these equations $\epsilon_{abc}$ denotes the fully antisymmetric three dimensional Levi-Civita symbol for which we shall use $\epsilon_{123}:=+1$. The vectors $(\bfl{e}_k)$, $k=1,2,3$ stand for the canonical basis of the $\mathbb{R}^3$, i.e., $\bfl{e}_1=(1,0,0)^T$ etc. The spin-moments $J_L$ are Cartesian STF (Symmetric and Trace-free) tensors, where $L$ is a Cartesian multi-index, $L = (k_1, \dots ,k_l)$ of $l$ Cartesian indices, each taking the values $1,2,3$. Symmetric refers to the symmetry with respect to all $l$ indices while trace-free means that every contraction between two arbitray indices vanishes. For more information on STF-tensors please also see \citet{thorne1980} and \citet{soffel:1994:139}. 

A summation over two equal (dummy) indices is implied automatically (e.g., $A_a B_a := A_1 B_1 + A_2 B_2 + A_3 B_3$) and $\partial_{ab} := \partial^2 /(\partial x^a \partial
x^b)$ etc. The hat on top of a symbol indicates the STF-part. Angle brackets indicate the STF-part with respect to the indices enclosed.

Note that for the first post-Newtonian metric the spin multipole moments  have to be defined to Newtonian order only. The spin-dipole moment, $\mathbf{J}$, agrees with the usual intrinsic angular momentum of
the body.

In this paper we will study a central axisymmetric body rotating uniformly about its symmetry axis. Perturbations of satellite orbits caused by the corresponding gravitomagnetic field induced by spin multipole moments of arbitrary order will be considered by means of perturbation theory. Such spin multipole moments have been considered by \citet{teyssandier1978} and \citet{panhans:2014paper}. Earlier applications of the mentioned expansions can be found in \citet{letelier2008} for the precession of a gyroscope and in \citet{iorio2001} where an alternative derivation of the Lense-Thirring effect is given. 

Here we shall also use a coordinate system with an axial symmetry with respect to the z-axis, i.e., the spin vector of the body points into z-direction. This simplifies the form of the used multipole expansion. On the other hand this might be a shortcoming because of the loss of generality due to unknown transformation behaviour of the spin moments. A treatment of a general spin orientation can be found in \citet{iorio2012} where a general spin-spin interaction was taken into account as well. A discussion about mixed effects connected with an arbitrary spin orientation can be found in \citet{iorio2015}.

\section{Equations of motion and STW-decomposition}
Post-Newtonian satellite equations of motion have been studied in detail in the literature (e.g., \citet{soffeldamourxu:1994}). Considering only a single isolated central body the coordinate acceleration of a satellite $\bfl{a}_{\text{S}}$ (e.g., equation (3.4) of \citet{soffeldamourxu:1994}) has the form
\begin{equation}
	\bfl{a}_{\text{S}} = \nabla \phi + \frac{1}{c^2} \left[ - 4 \phi \nabla \phi - \mathbf{v} (3 \dot{\phi} + 4 (\mathbf{v} \nabla) \phi) + 4 \mathbf{\dot{w}}  + v^{2} \nabla \phi + \mathbf{v} \times \mathbf{B} \right]
\end{equation}
where the gravitomagnetic field $\bfl{B}$ is defined by
\begin{equation}
	\bfl{B} := - 4 \nabla \times \bfl{w} \, .
\end{equation}
Since our central interest are the perturbations induced by some stationary gravitomagnetic field, i.e., $\dot{\mathbf{w}}=0$, we will simplify this equation to
\begin{equation}
	\bfl{a}_{\text{S}} = \nabla \phi + \bfl{a}_{\text{per}}
\end{equation}
with the perturbing acceleration
\begin{equation}
	\bfl{a}_{\text{per}} := \frac{1}{c^2} \bfl{v} \times \bfl{B} \, .
\end{equation}
Here, $\bfl{v}$ is the satellite's coordinate velocity. Effects connected with a variation in time of the spin vector and orders of magnitude were studied in \citet{mashoon2008} and \citet{iorio2002}. For satellite orbits around the Earth such time dependencies can be neglected.  

As mentioned in the beginning we will take $\phi = GM/r$ so the gravitoelectric part of the potential leads to unperturbed Keplerian orbits. The satellite orbit will be described by the usual set of orbital elements $(a,e,I,\Omega,\omega,M_0)$ and the perturbations by means of a Gaussian STW-perturbation theory of first order
with $S$, $T$, and $W$ being the scalar products of $\bfl{a}_{\text{per}}$ with $\bfl{n}$ for $S$, $\bfl{k}$ for $W$ and $\bfl{n} \times \bfl{k} $ for T. These vectors are defined by $\bfl{n} \coloneqq \bfl{x} / r$, $r:=\|x\|$ and $\bfl{k}:= \bfl{C}/C$, $\bfl{C} := \bfl{x} \times \bfl{v}$, $C := \|\bfl{C}\|$. 
 
An explicit calculation yields
\begin{equation}
	S = \frac{\mathbf{B} \cdot \mathbf{C}}{rc^2} , \quad
	T = -\frac{\mathbf{x} \cdot \mathbf{v}}{C} S, \quad 
	W = \frac{[(\mathbf{x}\mathbf{v})(\mathbf{v}\mathbf{B})-v^2(\mathbf{x}\mathbf{B})]}{Cc^2}. 
	\label{STW}
\end{equation} 
In a next step we want to calculate $\mathbf{B}$ explicitly.  Using \eqref{vectorpotential} a calculation of the $\mathbf{B}$-field reveals
\begin{equation}
	\mathbf{B} = 4 G \sum_{l=1}^\infty \frac{(-1)^l l}{(l+1)!} J_L \nabla  \partial_L \left( \frac{1}{r} \right) = 4 G \sum_{l=1}^{\infty} \frac{l(2l-1)!!}{(l+1)!} \nabla \left(\frac{J_L \hat{n}_L}{r^{l+1}} \right).
\end{equation} 
The use of spherical coordinates has proven to be of advantage and so we will use spherical spin multipole moments
\begin{equation}
	\Xi_{lm}:= \int d^3x \left[r^l (\mathbf{x} \times \boldsymbol{\sigma}) \nabla Y^*_{lm} \right]
\end{equation}
as introduced by \citet{panhans:2014paper}. Here 
\begin{equation}
	Y_{lm}(\lambda, \phi):=\sqrt{\frac{2l+1}{4 \pi} \frac{(l-m)!}{(l+m)!}} \text{e}^{im \phi} P_{lm}(\lambda)
\end{equation} 
stands for spherical harmonics and 
\begin{equation}
	P_{lm}(\lambda):=\frac{(-1)^m}{2^l l!} (1-\lambda^2)^{\frac{m}{2}} \frac{d^{m+l}}{d\lambda^{m+l}} (\lambda^2-1)^l
\end{equation}
are associated Legendre functions. Legendre polynomials are denoted as 
\begin{equation}
	P_l(\lambda):=P_{l0}(\lambda).
\end{equation} 
In the following the arguments of the these functions will be the polar angle $\varphi$ while we plug in $\cos(\theta)$ for $\lambda$ where $\theta$ is the azimuth angle. With this convention we will suppress arguments when using these functions. 

The spherical spin multipoles are connected with $J_L$ via
\begin{equation}
	J_L = \frac{4 \pi (l-1)!}{(2l+1)!!} \sum_{m=-l}^l \hat{Y}^{lm}_L \Xi_{lm}. \label{JL}
\end{equation}
Written in terms of spherical harmonics $\bfl{B}$ becomes
\begin{equation}
	\mathbf{B} = 4 G \sum_{l=1}^{\infty} \sum_{m=-l}^l \frac{4 \pi}{(2l+1)(l+1)} \Xi_{lm} \left\{ \frac{1}{r^{l+1}} \nabla Y_{lm} 
-(l+1) Y_{lm} \frac{\mathbf{x}}{r^{l+3}} \right\}. \label{bfinal2}
\end{equation}
The assumption of an axial symmetry implies $\Xi_{lm} = \xi_l \delta_{0m}$. Under this assumption expression \eqref{bfinal2} reduces to
\begin{equation}
	\mathbf{B} = - 4 G \sum_{l=1}^{\infty} \frac{2 \sqrt{\pi}}{\sqrt{(2l+1)}(l+1)} \xi_{l} \left\{ \frac{\sin(\theta) P_l'}{r^{l+2}} \mathbf{e}_{\theta}+ (l+1) P_l \frac{\mathbf{x}}{r^{l+3}} \right\},
\end{equation}
with
\begin{equation}
	\bfl{e}_{\theta} = \cos(\theta) \cos(\phi) \bfl{e}_1 + \cos(\theta) \sin(\phi) \bfl{e}_2 - \sin(\theta) \bfl{e}_3
\end{equation}
while $P_l'$ is the first derivative of $P_l$ with respect to its variable. 

As consequence of this form of $\bfl{B}$ we get for $S$ and $W$ ($T$ is given by $S$ through \eqref{STW})
\begin{align}	
	S = \; & \frac{8 \sqrt{\pi} C G}{c^2} \sum_{l=1}^{\infty} \frac{\xi_l}{\sqrt{2l+1}(l+1) r^{l+3}} \cos(I) P_l', \label{Sfinal2} \\  
	W = \; & \frac{8 \sqrt{\pi} C G}{c^2} \sum_{l=1}^{\infty} \frac{\xi_l}{\sqrt{2l+1}(l+1) r^{l+3}}  \nonumber \\ 
	& \left[ (l+1)P_l + \frac{re \sin(\nu) \cos(u) \sin(I)}{p} P_l' \right].  \label{Wfinal2}	   
\end{align} 
In this equation $\nu$ notes the true anomaly and we used $u:=\nu + \omega$ and $p:=a(1-e^2)$. In the next section the perturbations of the orbital elements will be discussed. 
\section{Discussion of the single orbital elements}
\label{sec:3}

\subsection{Semi-major axis}

The differential equation for the semi-major axis reads
\begin{equation}
	\dot{a} =\frac{2}{n \sqrt{1-e^2}} \left( Se \sin(\nu)+T \frac{p}{r} \right). 
\end{equation}
We make use of the second equation in \eqref{STW} and get 
\begin{equation}
	\dot{a} = \frac{2}{n \sqrt{1-e^2}} \left( e \sin(\nu) - \frac{\bfl{x} \cdot \bfl{v} }{C} \frac{p}{r} \right) S. 
\end{equation}
In the last step we furthermore apply
\begin{equation}
	\mathbf{x} \cdot \mathbf{v} = \frac{rC}{p}e \sin (\nu),
\end{equation} 
an equation we will also need for the eccentricity, and find
\begin{equation}
	\dot{a}=0. 
\end{equation}
So for this special form of the perturbation acceleration the semi-major axis will not change over time. 

\subsection{Eccentricity}
\label{subsecDiscus:1}

We are using again the second relation of \eqref{STW} and the expression for $\bfl{x} \cdot \bfl{v}$ together with equations known from the description with orbital elements,
\begin{equation}
	\cos(E) = \frac{r}{p}(e+\cos(\nu)), \quad \frac{p}{r} = 1+e \cos (\nu),
\end{equation}
and find
\begin{align}
\dot{e} = \frac{8 \sqrt{\pi} G \cos(I)}{c^2} \sum_{l=1}^{\infty} \frac{\xi_l}{\sqrt{2l+1}(l+1)a^{l+2}} \left( \frac{r}{a} \right)^{-1-l} \sin(\nu) P_l' \big( \sin(u) \sin(I) \big). 
\end{align}
The last three steps are expressing the $P_l'$ with $P_k$ via 
\begin{align}
	P_l' = \sum_{k=0}^{l-1} N_{lk} P_k, \quad N_{lk} = \begin{cases} 2k+1 & \quad \text{$k$ even, $l$ odd or $k$ odd, $l$ even} \\ 0 & \quad \text{otherwise} \end{cases},
\end{align}
(a consequence of no. 8.915/2. in \citet{gradstein:2007}) followed by rewriting the $P_k$ in terms of complex inclination functions $F_{kab}$ (see \citet{kaula1961})
\begin{equation}
	P_k(\sin(u) \sin(I)) = \sum_{b=0}^k F_{k0b}(I) \text{e}^{i(k-2b)u} 
\end{equation}
and finishing the series of conversions by eliminating the implicit time dependency through $\nu$ by using Hansen coefficients $X^{n,m}_s$ (e.g.,  \citet{hansennumerics:1990}) so we end with 
\begin{align}
	\dot{e} = & \; \frac{8 \sqrt{\pi} \cos (I) G}{c^2} \sum_{l=1}^{\infty} \frac{\xi_l}{\sqrt{2l+1}(l+1)a^{l+2}} \sum_{k=0}^{l-1} \sum_{b=0}^{k} N_{lk} F_{k0b}(I) \text{e}^{i \omega(k-2b)} \nonumber \\ 
	& \frac{1}{2i} \sum_{s=-\infty}^{\infty} \Big( X^{-1-l,k-2b+1}_s - X^{-1-l,k-2b-1}_s \Big) \text{e}^{isM}. \label{dote}
\end{align} 
In a first order perturbation theory we just have to integrate this expression with respect to $t$ which is fairly easy seeing the trivial time dependency $M(t)=nt+M_0$. One has to be careful with the term for $s=0$ though. The result is
\begin{align}
	\Delta e = & \; \frac{8 \sqrt{\pi} \cos (I) G}{c^2} \sum_{l=1}^{\infty} \frac{\xi_l}{\sqrt{2l+1}(l+1)a^{l+2}} \sum_{k=0}^{l-1} \sum_{b=0}^{k} N_{lk} F_{k0b}(I) \text{e}^{i \omega(k-2b)} \nonumber \\ 
	& \bigg\{ \frac{- 1}{2n} \sum_{\substack{s=-\infty \\ s \neq 0}}^{\infty} \Big( X^{-1-l,k-2b+1}_s - X^{-1-l,k-2b-1}_s \Big) \frac{\text{e}^{isM}}{s} \nonumber  \\
	& \hspace{0.3cm}+ \frac{1}{2i} \left( X^{-1-l,k-2b+1}_0 - X^{-1-l,k-2b-1}_0 \right)t \bigg\} \label{deltaet}.
\end{align}
The secular perturbations of $e$ vanish because of 
\begin{align}
	\Delta e_{\text{sec}}(t)  & = \sum_{l=1}^{\infty} \frac{8 t G \sqrt{\pi} \cos (I)  \xi_{l}}{\sqrt{2l+1}(l+1) a^{l+2}c^2} \sum_{\substack{k=0 \\ k=\text{even}}}^{l-1} \frac{N_{lk} F_{k0 \frac{k}{2}} (I)}{2i} \left( X^{-1-l,1}_0 - X^{-1-l,-1}_0 \right) \nonumber \\ 
	& = 0
\end{align}
where we used $X^{n,m}_0 = X^{n,-m}_0$. 
\subsection{Inclination}
\label{subsecDiscus:2}

The discussion of the inclination $I$ will follow the same pattern as before but gets slightly more complicated because of the appearance of $W$ rather than $S$. The starting point is the perturbation equation 
\begin{align}
	\dot{I} = \frac{8 \sqrt{\pi} G}{c^2} \sum_{l=1}^{\infty} \frac{\xi_l}{\sqrt{2l+1}(l+1)r^{l+2}} \cos(u) \left[ (l+1)P_l + \frac{r}{p}e \sin(\nu) \sin(I) \cos(u) P'_l\right].
\end{align}
This time we pass on expanding $P_l'$ but use 
\begin{equation}
	\cos(u) \sin(I) P_l'(\sin(u) \sin(I)) = \frac{\partial P_l}{\partial u}(\sin(u) \sin(I))
\end{equation}
instead. We apply the other conversions as before and find
\begin{align}
	\dot{I} = \; & \frac{4 \sqrt{\pi} G}{c^2} \sum_{l=1}^{\infty} \frac{\xi_l}{\sqrt{2l+1}(l+1)a^{l+1} p} \left( \frac{r}{a} \right)^{-1-l}  \nonumber \\ 
	& \sum_{b=0}^l F_{l0b} \left( \text{e}^{iu}+\text{e}^{-iu} \right) \text{e}^{iu(l-2b)} \Big[ (l+1)(1+e \cos(\nu)) + i e (l-2b) \sin(\nu) \Big] \nonumber \\
\end{align}
\begin{align}	
	= \; & \frac{4 \sqrt{\pi} G}{c^2} \sum_{l=1}^{\infty} \frac{\xi_l}{\sqrt{2l+1}(l+1)a^{l+1} p} \left( \frac{r}{a} \right)^{-1-l} \sum_{b=0}^l F_{l0b} \text{e}^{i \omega(l-2b)} \nonumber \\ 
	& \bigg\{  \left[ (l+1) \text{e}^{i \nu(l-2b+1)} + \frac{e}{2} (2l-2b+1) \text{e}^{i \nu(l-2b+2)} +\frac{e}{2} (2b+1) \text{e}^{i \nu(l-2b)} \right] \text{e}^{i \omega} \nonumber \\ 
	& + \left[ (l+1) \text{e}^{i \nu(l-2b-1)} + \frac{e}{2} (2l-2b+1) \text{e}^{i \nu(l-2b)} +\frac{e}{2} (2b+1) \text{e}^{i \nu(l-2b-2)} \right] \text{e}^{-i \omega} \bigg\} \nonumber \\
	= & \; \sum_{l=1}^{\infty} \frac{ 4 \sqrt{\pi} G \xi_l}{\sqrt{2l+1}(l+1)a^{l+2} (1-e^2)c^2} \sum_{b=0}^l F_{l0b} \text{e}^{i \omega(l-2b)} \nonumber \\ 
	& \sum_{s=-\infty}^{\infty} \Bigg\{ \left[ (l+1) X^{-1-l,l-2b+1}_s + \frac{e}{2} (2l-2b+1) X^{-1-l,l-2b+2}_s \right.  \nonumber \\ 
	& \hspace{1cm} \left. + \frac{e}{2} (2b+1) X^{-1-l,l-2b}_s \right] \text{e}^{i \omega} + \Big[ (l+1) X^{-1-l,l-2b-1}_s \nonumber \\ 
	&  \hspace{1cm} + \frac{e}{2} (2l-2b+1) X^{-1-l,l-2b}_s +\frac{e}{2} (2b+1) X^{-1-l,l-2b-2}_s \Big] \text{e}^{-i \omega} \Bigg\} \text{e}^{isM}.
\end{align}
An integration yields the fairly long expression
\begin{align}
	\Delta I(t) = \; &  \sum_{l=1}^{\infty} \frac{4 G \sqrt{\pi} \xi_l}{\sqrt{2l+1}(l+1)a^{l+2} (1-e^2)c^2} \sum_{b=0}^l F_{l0b} \text{e}^{i \omega(l-2b)} \nonumber \\ 
	& \Bigg\{ \sum_{\substack{s=-\infty \\ s \neq 0}}^{\infty} \bigg\{ \left[ (l+1) X^{-1-l,l-2b+1}_s + \frac{e}{2} (2l-2b+1) X^{-1-l,l-2b+2}_s \right. \nonumber \\ 
	&  \hspace{0.5cm}  + \left. \frac{e}{2} (2b+1) X^{-1-l,l-2b}_s \right] \text{e}^{i \omega} + \Big[ (l+1) X^{-1-l,l-2b-1}_s \nonumber \\ 
	&  \hspace{0.5cm} + \left. \frac{e}{2} (2l-2b+1) X^{-1-l,l-2b}_s +\frac{e}{2} (2b+1) X^{-1-l,l-2b-2}_s \right] \text{e}^{-i \omega} \bigg\} \frac{ \text{e}^{isM}}{i n s} \nonumber \\
	&  \hspace{0.5cm} + \bigg\{ \left[ (l+1) X^{-1-l,l-2b+1}_0 + \frac{e}{2} (2l-2b+1) X^{-1-l,l-2b+2}_0  \right. \nonumber \\ 
	&  \hspace{0.5cm} + \frac{e}{2} \left. (2b+1) X^{-1-l,l-2b}_0 	\right] \text{e}^{i \omega} + \Big[ (l+1) X^{-1-l,l-2b-1}_0 \nonumber \\ 
	& \hspace{0.5cm} + \frac{e}{2} (2l-2b+1) X^{-1-l,l-2b}_0 +\frac{e}{2} (2b+1) X^{-1-l,l-2b-2}_0 \Big] \text{e}^{-i \omega} \bigg\} t \Bigg\}. \label{deltaihansen}
\end{align}
As for the secular perturbation we find
\begin{align}
	\Delta I_{\text{sec}}(t) = & \;  \frac{4 \sqrt{\pi} G t}{c^2} \sum_{\substack{l=1 \\ l=\text{odd}}}^{\infty} \frac{\xi_l}{\sqrt{2l+1}(l+1) a^{l+2} (1-e^2)} \nonumber \\
	& \; \left\{ F_{l0 \frac{l+1}{2}} \left[ (l+1) X^{-1-l,0}_0 + \frac{e}{2} l X^{-1-l,1}_0 + \frac{e}{2} (l+2) X^{-1-l,-1}_0 \right] \right. \nonumber \\
	&  \left. + F_{l0 \frac{l-1}{2}} \left[ (l+1) X^{-1-l,0}_0 + \frac{e}{2} (l+2) X^{-1-l,1}_0 + \frac{e}{2} l X^{-1-l,-1}_0 \right] \right\} \nonumber \\
	= & \;  \frac{4 \sqrt{\pi} G t}{c^2} \sum_{l=1}^{\infty} \frac{\xi_l}{\sqrt{2l+1}(l+1) a^{l+2} (1-e^2)} \nonumber \\
	& \; \left( F_{l0 \frac{l+1}{2}} + F_{l0 \frac{l-1}{2}} \right) \Big( l+1 \Big) \left( X^{-1-l,0}_0 + e  X^{-1-l,1}_0 \right) \nonumber \\
	= & \; 0,
\end{align}
which vanishes again due to $F_{l0 \frac{l+1}{2}} + F_{l0 \frac{l-1}{2}} = 0 $ for odd numbers $l$. 
\subsection{Argument of the ascending node}
\label{subsecDiscus:3}
Luckily the above argumentation can be adopted for the calculation of $\Delta \Omega$ with some minor changes since the differential equations for $\Omega$ and $I$ just differ by the appearance of $\sin(u)$ rather than $\cos(u)$ and an additional $\sin(I)$ in the denominator which has no influence on the calculation at all. 
So the result for $\Delta \Omega$ reads
\begin{align}
	\Delta \Omega(t) = \; &  \sum_{l=1}^{\infty} \frac{-i4 G \sqrt{\pi} \xi_l}{\sin(I) \sqrt{2l+1}(l+1)a^{l+2} (1-e^2) c^2} \sum_{b=0}^l F_{l0b} \text{e}^{i \omega(l-2b)} \nonumber \\ 
	& \Bigg\{ \sum_{\substack{s=-\infty \\ s \neq 0}}^{\infty} \bigg\{ \left[ (l+1) X^{-1-l,l-2b+1}_s + \frac{e}{2} (2l-2b+1) X^{-1-l,l-2b+2}_s  \right.  \nonumber \\ 
	& \hspace{0.5cm} + \left. \frac{e}{2} (2b+1) X^{-1-l,l-2b}_s \right] \text{e}^{i \omega} - \Big[ (l+1) X^{-1-l,l-2b-1}_s \nonumber \\ 
	& \hspace{0.5cm} + \left. \frac{e}{2} (2l-2b+1) X^{-1-l,l-2b}_s +\frac{e}{2} (2b+1) X^{-1-l,l-2b-2}_s \right] \text{e}^{-i \omega} \bigg\} \frac{\text{e}^{isM}}{i n s} \nonumber \\
	& \hspace{0.5cm}  + \bigg\{ \left[ (l+1) X^{-1-l,l-2b+1}_0 + \frac{e}{2} (2l-2b+1) X^{-1-l,l-2b+2}_0   \right. \nonumber \\ 
	& \hspace{0.5cm} + \left. \frac{e}{2} (2b+1) X^{-1-l,l-2b}_0 \right] \text{e}^{i \omega} - \Big[ (l+1) X^{-1-l,l-2b-1}_0  \nonumber \\ 
	& \hspace{0.5cm} + \frac{e}{2} (2l-2b+1) X^{-1-l,l-2b}_0 +\frac{e}{2} (2b+1) X^{-1-l,l-2b-2}_0 \Big] \text{e}^{-i \omega} \bigg\} t \Bigg\}. \label{deltaOmega}
\end{align}
This has noticeable consequences in particular for the secular perturbations which read 
\begin{align}
	\Delta \Omega_{\text{sec}}(t) = & \; \frac{8 \sqrt{\pi} G t}{ c^2 \sin(I)} \sum_{\substack{l=1 \\ l=\text{odd}}}^{\infty} \frac{\xi_l}{\sqrt{2l+1}a^{l+2}} \Im \left( F_{l0\frac{l+1}{2}} \right) X^{-2-l, 0}_0 \nonumber \\
	= & \; \frac{8 \sqrt{\pi} G t}{c^2 \sin(I)}  \sum_{\substack{l=1 \\ l=\text{odd}}}^{\infty} \frac{\xi_{l}}{\sqrt{2l+1} a^{l+2} \left(1-e^2 \right)^{l+\frac{1}{2}}} \Im \left( F_{l0\frac{l+1}{2}}\right)  \nonumber \\ 
	& \sum_{n=0}^{\frac{l-1}{2}} \left( \frac{e}{2} \right)^{2n} \binom{2n}{n} \binom{l}{2n} \label{Omegasnu}
\end{align} 
where we used 
\begin{equation}
	(1-e^2)X^{n,m}_s = X^{n+1,m}_s + \frac{e}{2} \left( X^{n+1,m+1}_s + X^{n+1,m-1}_s \right)
\end{equation}
with $s=m=0$, $n=-l-2$ and
\begin{equation}
	X^{-(n+1),m}_0 = \left( \frac{e}{2} \right)^m \frac{1}{(1-e^2)^{n-\frac{1}{2}}} \sum_{b=0}^{\left[ \frac{n-m-1}{2}\right]} \left( \frac{e}{2} \right)^{2b} \binom{2b+m}{b} \binom{n-1}{2b+m}
\end{equation}
for $n \in \mathbb{N}$, $m \in \mathbb{Z}$.

If only the spin-dipole is kept we recover the well known result of Lense and Thirring (\citet{lense:1984}):
\begin{equation}
	\Delta \Omega_{\text{sec}}^{l=1} = \frac{8 \sqrt{\pi} Gt}{c^2 \sin(I)} \frac{\xi_1}{\sqrt{3}a^3(1-e^2)^{\frac{3}{2}}} \Im \left( F_{101} \right) = \frac{2G}{c^2 a^3 n (1-e^2)^{\frac{3}{2}}} \sqrt{\frac{4\pi}{3}} \xi_1 nt.  
\end{equation}
Inserting $\xi_1 = \sqrt{3/4\pi}J$ the Lense-Thirring expression is obtained. 

Because of its relevance we want to apply this formula to the LAGEOS 2 satellite. We will treat Earth as a homogeneous body. Since the quadrupole term does not contribute to the secular perturbations at all the dipole needs to be compared to the octupole term. Evaluating formula \eqref{Omegasnu} gives
\begin{align}
\dot{\Omega}_{\text{sec}}^{l=3} = & \; 3 \sqrt{\frac{\pi}{7}} \frac{G \xi_3 (1+\frac{3}{2}e^2)}{c^2 a^5 (1-e^2)^{\frac{7}{2}}} \nonumber \\ & \left( 4 \cos^5 \left( \frac{I}{2} \right) \sin \left( \frac{I}{2} \right) - 12 \cos^3 \left( \frac{I}{2} \right) \sin^3 \left( \frac{I}{2} \right) + 4 \cos \left( \frac{I}{2} \right) \sin^5 \left( \frac{I}{2} \right) \right)
\end{align} 
for the secular drift rate caused by the spin octupole. For the  model of a homogeneous Earth one finds 
\begin{equation}
\dot{\Omega}^{l=3}_{\text{sec}} = 0.02 \frac{\text{mas}}{\text{yr}}.
\end{equation}
Since our model of a homogeneous Earth overestimates the value of the spin octupole moment the actual value for the drift rate is even smaller than the calculated number. If one compares this value to the spin dipole term (\citet{ciufolini:2004}),
\begin{equation}
\dot{\Omega}^{l=1}_{\text{sec}} = 31.5 \frac{\text{mas}}{\text{yr}},
\end{equation}
it becomes obvious that the contribution of the spin octupole and all higher spin multipole moments can be neglected for Earth's satellite orbits at present.  
  
\subsection{Argument of periapsis}
\label{subsecDiscus:4}
Most of the work for this orbital element is done by now because the result from \ref{subsecDiscus:3} can be used. However, in the differential equation for $\omega$ appears an additional term $\dot{\omega}_{\text{add}}$ which needs to be studied. So we calculate
\begin{align}
	\dot{\omega}_{\text{add}} := & \; \frac{\sqrt{1-e^2}}{nae} 
	\Bigg\{ -S \cos(\nu)+T \sin(\nu)\left[1+\frac{r}{p}\right] \Bigg\} \nonumber \\ 
	= & \; - \frac{C}{\mu e} \left[ \cos(\nu) + \frac{r}{p}e \sin^2(\nu) + \left( \frac{r}{p} \right)^2 e \sin^2(\nu)\right]S \nonumber \\ 
	= & \; - \frac{8 \sqrt{\pi} p G \cos(I)}{e c^2} \sum_{l=1}^{\infty} \frac{\xi_l}{\sqrt{2l+1}(l+1)r^{l+3}} \nonumber \\ 
	& \; \left( \cos(\nu) + \frac{r}{p}e \sin^2(\nu) + \left( \frac{r}{p} \right)^2 e \sin^2(\nu) \right) \sum_{k=0}^{l-1} \sum_{b=0}^k N_{lk} F_{k0b}(I) \text{e}^{iu(k-2b)}  
\end{align}
\begin{align}	
	= & \; - \frac{8 \sqrt{\pi} p G \cos(I)}{e c^2} \sum_{l=1}^{\infty} \frac{\xi_l}{\sqrt{2l+1}(l+1)a^{l+3}} 
	\sum_{k=0}^{l-1} \sum_{b=0}^k N_{lk} F_{k0b}(I) \text{e}^{i\omega(k-2b)} \nonumber \\ 
	& \; \sum_{s=-\infty}^{\infty} \Bigg\{ \frac{1}{2} \left( X^{-l-3,k-2b+1}_s + X^{-l-3,k-2b-1}_s \right) \nonumber \\ 
	&  \hspace{0.5cm} + \frac{e}{2(1-e^2)} \left[ X^{-l-2,k-2b}_s - \frac{1}{2} \left( X^{-l-2,k-2b+2}_s + X^{-l-2,k-2b-2}_s \right) \right] \nonumber \\
	& \hspace{0.5cm} + \frac{e}{2(1-e^2)^2}  \left[ X^{-l-1,k-2b}_s - \frac{1}{2} \left( X^{-l-1,k-2b+2}_s + X^{-l-1,k-2b-2}_s \right) \right] \Bigg\} \text{e}^{isM}. \label{dotomega_2}
\end{align}
We integrate \eqref{dotomega_2} and find 
\begin{align}
	\Delta \omega_{\text{add}}(t) & = - \sum_{l=1}^{\infty} \frac{8 \sqrt{\pi} G p \cos(I) \xi_l}{e c^2 \sqrt{2l+1}(l+1)a^{l+3}} 
	\sum_{k=0}^{l-1} \sum_{b=0}^k N_{lk} F_{k0b}(I) \text{e}^{i\omega(k-2b)} \nonumber \\ 
	\Bigg\{ & \sum_{\stackrel{s=-\infty}{s \neq 0}}^{\infty} \bigg\{ \frac{1}{2} \left( X^{-l-3,k-2b+1}_s + X^{-l-3,k-2b-1}_s \right) \nonumber \\ 
	+ & \frac{e}{2(1-e^2)} \left[ X^{-l-2,k-2b}_s - \frac{1}{2} \left( X^{-l-2,k-2b+2}_s + X^{-l-2,k-2b-2}_s \right) \right] \nonumber \\
	+ & \frac{e}{2(1-e^2)^2}  \left[ X^{-l-1,k-2b}_s - \frac{1}{2} \left( X^{-l-1,k-2b+2}_s + X^{-l-1,k-2b-2}_s \right) \right] \bigg\} \frac{\text{e}^{isM} }{ins} \nonumber \\
	+ & \bigg\{ \frac{1}{2} \left( X^{-l-3,k-2b+1}_0 + X^{-l-3,k-2b-1}_0 \right) \nonumber \\ 
	+ & \frac{e}{2(1-e^2)} \left[ X^{-l-2,k-2b}_0 - \frac{1}{2} \left( X^{-l-2,k-2b+2}_0 + X^{-l-2,k-2b-2}_0 \right) \right] \nonumber \\
	+ &\frac{e}{2(1-e^2)^2}  \left[ X^{-l-1,k-2b}_0 - \frac{1}{2} \left( X^{-l-1,k-2b+2}_0 + X^{-l-1,k-2b-2}_0 \right) \right] \bigg\}t \Bigg\}. \label{deltaomega_2}
\end{align}
So using the result from \ref{subsecDiscus:3} we find for the overall perturbation 
\[
	\Delta \omega (t) = -\cos(I) \Delta \Omega(t) + \Delta \omega_{\text{add}}(t)
\]
with $\Delta \Omega(t)$ from \eqref{deltaOmega} and $\Delta \omega_{\text{add}}(t)$ from \eqref{deltaomega_2}. The secular perturbations are given by 
\begin{align}
	\Delta \omega_{\text{sec}}(t) = \; &  -\cos{I} \Delta \Omega_{\text{sec}}(t) - \sum_{l=1}^{\infty} \frac{8 \sqrt{\pi} G p \cos(I) t \xi_l}{e c^2 \sqrt{2l+1}(l+1)a^{l+3}} 
	\sum_{\substack{k=0 \\ k=\text{even}}}^{l-1} N_{lk} F_{k0\frac{k}{2}} \nonumber \\ 
	& \Bigg\{ X^{-l-3,1}_0 + \frac{e\left( X^{-l-2,0}_0 - X^{-l-2,2}_0 \right)}{2(1-e^2)} + \frac{e\left( X^{-l-1,0}_0 - X^{-l-1,2}_0 \right)}{2(1-e^2)^2}   \Bigg\} \nonumber \\
\end{align}	

Again, for the spin-dipole the result agrees with the classical expression of Lense and Thirring:
\begin{align}
	\omega_{\text{sec}}^{l=1} = & \; -\cos(I) \Delta \Omega_{\text{sec}}^{l=1} - \frac{8 \sqrt{\pi} G p \cos(I)}{e c^2} \frac{\xi_1}{\sqrt{3}a^{4}} \frac{e}{(1-e^2)^{\frac{5}{2}}} t \nonumber \\ 
	= & \; -3 \cos(I) \Delta \Omega_{\text{sec}}^{l=1}.
\end{align}

\section{Summary}
\label{sec:4}
The aim of the paper was to investigate the influence of a central gravity field with a mass monopole and arbitrary spin multipole moments on satellite orbits given a stationary axisymmetric setting. In order to simplify the form of the multipole moments a coordinate system was chosen for which the spin vector pointed into the z-direction. We found that $\dot{a}=0$ holds in general (and not just in first order perturbation theory) and that $e$ and $I$ experience perturbations which have no secular contributions. For odd numbers $l$ there are secular perturbations for $\Omega$ and $\omega$ which yield for the spin-dipole, $l=1$, the well known results by Lense and Thirring. In the case of the LAGEOS 2 satellite we calculated the additional secular drift due to Earth's spin octupole and found a negligible small number. Further Physical implications and orders of magnitude will be discussed elsewhere.

%
%


\section*{Conflict of Interest}
The authors declare that they have no conflict of interest.

\bibliographystyle{apalike}
\bibliography{literature}   

\end{document}